\documentclass[sn-mathphys]{sn-jnl}

\jyear{2023}%

\theoremstyle{thmstyleone}%
\usepackage{verbatim}

\usepackage{setspace}
\usepackage{silence}
\usepackage{lmodern}
\usepackage{url}
\usepackage[export]{adjustbox}
\usepackage{longtable}

\usepackage[T1]{fontenc}
\usepackage{array}
\usepackage{enumitem}

\makeatletter
\expandafter\let\csname algorithm*\endcsname\relax
\expandafter\let\csname endalgorithm*\endcsname\relax
\makeatother
\usepackage[ruled,vlined]{algorithm2e}
\theoremstyle{thmstyleone}%

\usepackage{graphicx}
\usepackage{float}
\SetKwInput{KwInput}{Input}                
\SetKwInput{KwOutput}{Output} 
\usepackage{lineno}
\modulolinenumbers[5]
\usepackage{setspace}
\usepackage{amsmath,amsfonts,amssymb}
\usepackage{adjustbox}
\usepackage{listingsutf8}
\usepackage[utf8]{inputenc}

\begin{document}

\title{TUANDROMD-X: Advanced Entropy and Visual Analytics Dataset for Enhanced Malware Detection and Classification}

\author*[1]{\fnm{Parthajit} \sur{Borah}}\email{parthajit@tezu.ernet.in}

\author[1]{\fnm{Upasana Sarmah} }\email{upatink@tezu.ernet.in}

\author[1]{\fnm{D.K. Bhattacharyya} }\email{dkb@tezu.ernet.in}
\author[2]{\fnm{J.K. Kalita} }\email{jkalita@uccs.edu.in}

\affil*[1]{\orgdiv{Department
		of Computer Science and Engineering}, \orgname{Tezpur University}, \orgaddress{ \city{Tezpur}, \postcode{784028}, \state{Assam}, \country{India}}}
\affil[2]{\orgdiv{Computer Science, College of Engineering and Applied Science}, \orgname{University of
Colorado,Colorado Springs}, \orgaddress{ \postcode{CO 80933-7150}, \state{Colorado}, \country{USA}}}


\abstract{Malware and malware-based attacks are becoming more prevalent and complex.
Attackers regularly come up with new techniques that have the ability to evade
conventional and signature-based malware defense. In order to address
such threats, there is an increasing demand for advanced and better
defense solutions. Machine learning-based techniques are efficiently capable
of defending against malware and malware-based attacks. Nevertheless,
creating and efficiently testing such techniques demand high-quality datasets
having samples of various malware families as well as
goodware. The lack of such datasets continues to be a major bottleneck
in malware research. In this paper, we introduce TUANDROMD-X, a
multiclass malware dataset with visual and entropy-based features of each sample, distinctly identifying malware from goodware.
The dataset is created based on static analysis, lowering the overhead
that comes with high feature engineering and dynamic analysis.
As a result, TUANDROMD-X facilitates researchers and cybersecurity
experts to design faster and better malware detection
systems.

}

\keywords{Malware, Entropy, Static, Dynamic, Android}



\maketitle
\section{Introduction}
Advancements of digital technologies in different sectors demand high-end security mechanisms as these systems face significant threats from malware and malware-based
attacks. Primarily, Malware, malicious software is designed to orchestrate disruption, damage, or even gain unauthorized access to computer systems. Malware presents a multifaceted challenge due to its evolving nature and increasing sophistication. These malicious entities range from traditional viruses and worms to advanced ransomware and spyware, capable of executing complex attack vectors that can bypass conventional security measures.
Malware-based attacks exploit system vulnerabilities to perform unauthorized activities, such as data theft, espionage, and operational disruption, leading to substantial financial loss and damage to reputation.\\ 
Nowadays, malware authors employ a variety of sophisticated techniques, such as obfuscation, to evade detection systems \cite{o2011obfuscation}. Obfuscation methods can include packing and encryption. On the other hand, code obfuscation masks the true nature and intent of malicious software, making it challenging for traditional detection methods to identify them. Malware can be categorized based on the evasive techniques it uses to thwart detection systems such as encrypted, oligomorphic, polymorphic, and metamorphic malware \cite{you2010malware}. To counteract these advanced evasion tactics, defense mechanisms must be highly resilient and capable of detecting even the most cunningly concealed malware. Typically, static and dynamic analysis are performed to build such robust systems. Static analysis involves examining the code, structure, and static properties of executables without executing them, which can reveal certain patterns or signatures associated with malware \cite{moser2007limits}. However, static analysis can be thwarted by obfuscation techniques that disguise the code. Dynamic analysis, on the contrary, involves executing the code in a controlled environment to observe its behavior, which can detect actions indicative of malicious intent \cite{or2019dynamic}. Despite its effectiveness, dynamic analysis is resource-intensive, can be bypassed by malware designed to detect virtualized environments, and requires extensive feature engineering to interpret the behaviors accurately.\\ 
To combat these issues, there is a growing need for advanced malware data representation technique. These novel approaches aim to generate more efficient and compact representations of malware behaviors and structures, which can be processed more quickly and don't require extensive feature engineering. By developing such representations, we can enhance the speed and effectiveness of malware analysis, leading to more responsive and adaptive defense systems. This advancement would allow for real-time threat detection and response, even against rapidly evolving malware variants, thereby significantly improving overall cybersecurity postures. \\

\subsection{Contribution}
The primary contribution of this paper is the creation of two large datasets of malware, which are both referred to as TUANDROMD-X. TUANDROMD-X utilizes sophisticated static techniques, including entropy analysis and visual analytics, to create two large datasets. The two datasets are large and diverse, with each having 30,000 instances that are classified into 72 classes.
71 classes in total cover different kinds of malware, and one class to cover goodware. Such data classification aids in performing sophisticated analyses of a wide variety of malware types. In addition, TUANDROMD-X is made available under a CC-BY license. Such an open policy promotes collaboration, invites further development within the community, and allows for quicker development of better malware detection methods within the broader cybersecurity community. Researchers who want to use this dataset for further research can do so by requesting the same through a letter to the corresponding author via an email.

\subsection{Motivation}
The Internet is a powerhouse of resources as almost all industry, and organization are banking on it to create, grow and maintain their customer base. Ranging from financial, education, entertainment, gaming and gambling almost all services are provided over the Internet. Primarily, all Web applications exchange information to and from its users by relying on different protocols. Since, the past two decades malware attacks have served as a major security problem as attackers exploit vulnerabilities in the applications and the host systems. The sophistication of these attacks are troublesome not only for the users but also for the security experts and researchers. Defending against the attacks and shielding the users and services of these applications is a challenging task but not impossible. Over the years, numerous detection mechanisms have been developed but with these developments come another challenge  the evolving of new attack vectors. Smart attackers employ these vectors wisely to generate attack payload which has very similar characteristics with normal operations. A very significant aspect of developing a detection method is the data resources on which it is based. A quality dataset helps assess the biases of a method in terms of its performance and enhances the detection rates of a defense system ensuring its reliability.

\subsection{Organisation of the paper}\label{organisation}
The organisation of the paper is as follows. In Section \ref{background}, the background and state-of-the-art methods are discussed in length. Section \ref{methodology} on the other hand presents the methodology used to develop TUANDROMD-X starting from the data collection process, entropy analysis, and Greyscale data mapping. Section \ref{results} illustrates the results and experimentation using TUANDROMD-X. Lastly, we wind up with the overall concluding remarks in Section \ref{conclusion}. 
\section{Background and State of the Art}\label{background}
Malicious software, or malware, is any program intentionally designed to cause harm to computer systems or networks. These programs are crafted in such a way that they are able to infiltrate computer systems and network
resources. In doing so, the main aim is to obtain sensitive information and sabotage normal operations, in most cases without the knowledge or consent of the system’s owners. When this
happens widely, it create devastation that permeates the entire Internet.
Malicious software encountered in real-world scenarios manifests in diverse
forms. The prevalent types include Adware, Trojan, Backdoor, Worm, Bot, Rootkit, Downloader,
Ransomware, and Virus. Malware attacks can vary greatly in complexity and sophistication. Some
attacks are simple and involve only a single stage, while others are more complex and involve multiple stages and different types of malware. In simpler
attacks, a single malicious code or technique may be used to breach a system
or compromise security. In more complex attacks, a series of interconnected
steps may occur, each with a specific purpose in compromising the target. To effectively counter network intrusion events, we require knowledge of malware instances and their behaviors. This information is obtained by acquiring and analyzing the malware itself. While malware appears in many different forms, certain common techniques are used to analyze it, revealing the risks it poses and the intentions behind its creation. Malware analysis helps to identify the indicators of compromise, which are then used to build detection models. There are four fundamental approaches to malware analysis: static, dynamic, automated, and manual code reversing \cite{egele2008survey} \cite{chakkaravarthy2019survey}. Each of these methods provides valuable insights that contribute to the overall understanding and mitigation of malware threats.

In the realm of combating malware threats, the representation of malware data holds significant importance as it fortifies defense systems against a diverse array of malicious attacks. These representations not only help to detect but also mitigate malicious software invasions. Malware data is presented in several formats, each catering to unique aspects of threat analysis and defense mechanisms. These representations serve as input data for malware defense models and security measures. Malware data can be categorized into distinct formats, including tabular data, image data, graph data, sequence data, and text data. 

There are a handful of publicly known malware datasets like MalNet \cite{freitaslarge}, MalImage \cite{nataraj2011comparative}, and MalViz \cite{nguyen2018malviz}. The datasets comprise only grayscale malware images and are missing entropy analysis. Other feature datasets exist but are small and time-consuming to build. TUANDROMD-X, nonetheless, offers an end-to-end solution with 30,000 samples including grayscale images and entropy analysis. The dataset includes malware and goodware of 72 classes—71 for different types of malware families and 1 for goodware.

\section{Methods}\label{methodology}
\begin{figure}[ht!]
	\centering
\includegraphics[width=0.8\linewidth]{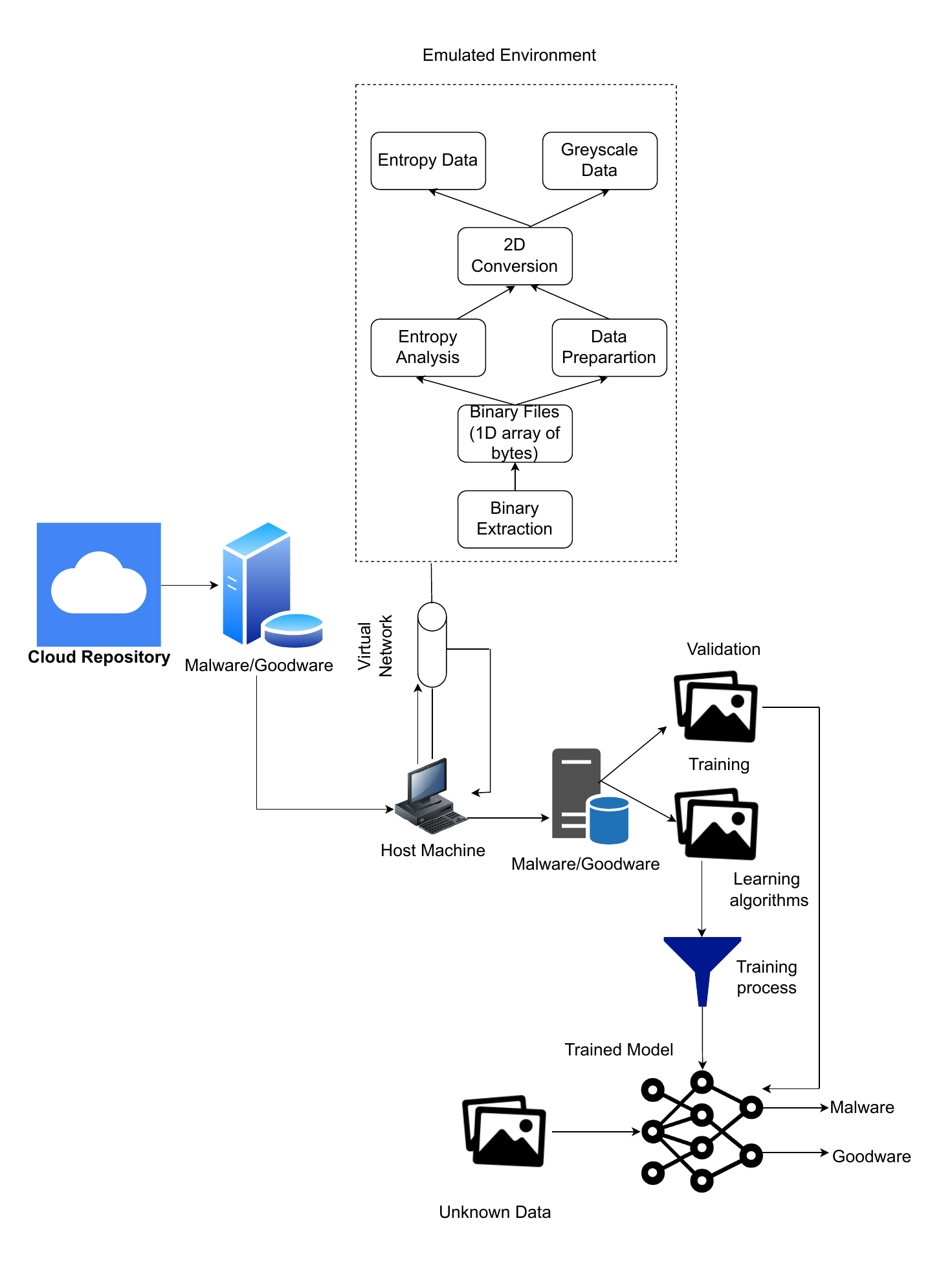}
	\caption{Proposed Framework}
	\label{figure:tundrod1}
\end{figure}

This section describes the process of generating TUANDROMD-X's multimodal nature by combining entropy and visual data of malware. The creation of TUANDROMD-X involves a meticulous procedure to ensure the integration of both entropy metrics and visual greyscale representations. The proposed framework of TUANDROMD-X, illustrated in Figure \ref{figure:tundrod1}, outlines the overall process of dataset creation and evaluation.
\subsection{Data Collection}
10,000 benign Android applications were collected from the Google Play Store and AndroZoo to create the goodware dataset. For malware binaries, we collected 20000 Android raw malware binaries from 71 families from \cite{wei2017deep} \cite{borah2020malware}. A database server is used to store all the android apps collected for further analysis and processing. This large collection is used as the basis for creating the dual-modal representations of the data, such as entropy metrics and visual greyscale images, to enhance malware detection and classification.

\subsection{Entropy Analysis}
\begin{figure}[ht!]
	\centering
	\includegraphics[width=1\linewidth]{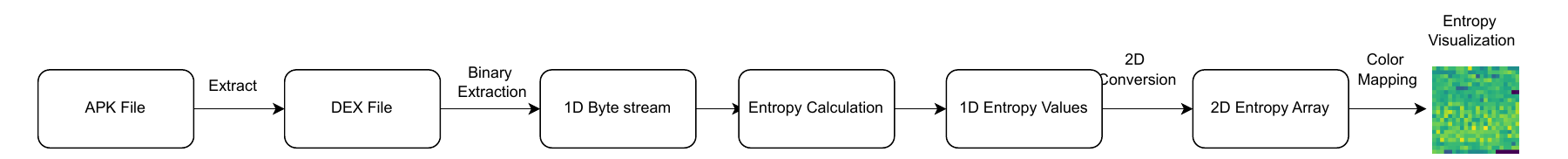}
	\caption{Steps of Entropy Analysis}
	\label{figure:tundrod2}
\end{figure}

The first step of performing entropy analysis of an APK file is to load the raw malware binary into a byte object, as shown in Figure \ref{figure:tundrod2}. The byte object is an array of values (integers between 0 and 255) that make up the raw data of the APK file. The following byte stream is the foundation for the computation of entropy. This first conversion to the byte sequence is necessary since it makes it possible to analyze the file content uniformly regardless of its original structure or format.\\
When the APK file is converted to this sequence of bytes, the correct entropy analysis begins. Entropy in information theory is applied to quantify statistical uncertainty or randomness in data. The analysis procedure involves a method known as the sliding window method. This is a method of scanning the binary data in constant-size windows, in this case 256 bytes; although this can be changed depending on the particular requirements of the analysis.
For each window of bytes, the following steps are performed:
\begin{enumerate}
    \item \textbf{Byte Frequency Calculation}: The frequency of each possible byte value (0 to 255) within the window is counted. This step creates a histogram of byte occurrences for the current segment.
    \item \textbf{Probability Distribution}: The frequency count is then normalized to create a probability distribution. Each probability in this distribution represents the likelihood of encountering a specific byte value within the current window.
    \item \textbf{Entropy Computation \cite{gray2011entropy}}: Using this probability distribution, Shannon entropy is computed, which quantifies the randomness of the data in that segment. The formula for Shannon entropy is:
    \begin{equation}
       H(X) = -\sum p(x) \log_2 p(x)
        \label{eq:shannon_entropy}
    \end{equation}
    where \( p_x \) is the probability of byte \( x \) occurring in the window, and the sum is taken over all byte values present in the window.
    \item \textbf{Window Sliding}: After calculating the entropy for one window, the process moves forward by a certain step size of 1 Window size (256 bytes) to create a new overlapping window, and the process repeats for all sliding windows across the entire byte stream of the binary file.
\end{enumerate}

The resulting entropy values provide a detailed view of the data's randomness across different file segments. After entropy value calculation, normalizing these values involves scaling the raw entropy measurements to fit within a standard range (0-255) using min-max normalization. This process ensures that the entropy values can accurately represent pixel intensities in an image. After normalization, the 1D array of normalized entropy measurements is converted into a 2D matrix suitable for image representation. A color map is applied to this matrix for visual representation of the entropy values, and the result is saved as an image file. This visualization technique aids in identifying patterns, anomalies, and structural characteristics within the binary data of an APK file.

\begin{figure}[ht!]
	\centering
	\includegraphics[width=0.6\linewidth]{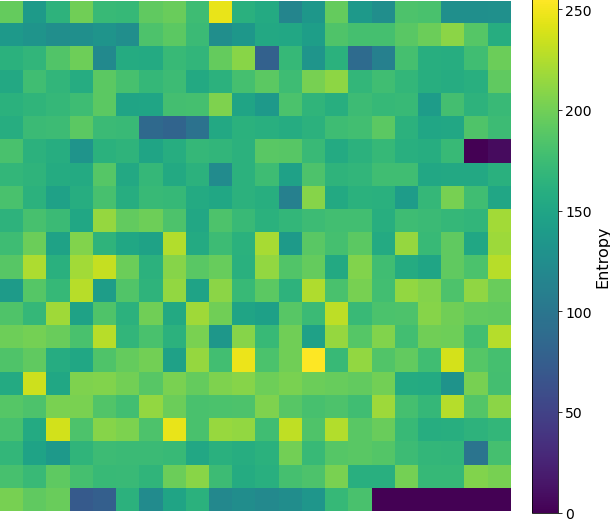}
	\caption{Entropy Visualisation of an application}
	\label{figure:tundrod3}
\end{figure}
Colormaps \cite{hunter2007matplotlib} are used to represent entropy values, indicating data randomness within files. Typically, a file is analyzed in small chunks, with entropy calculated for each. These values are then mapped to colors, often using a scale where dark purple represents low entropy and yellow represents high entropy. The low entropy area might represent padding, uninitialized data, or a resource section with uniform content. In an APK file, other areas that might have low entropy include the AndroidManifest.xml file, which contains metadata about the app and tends to have structured and predictable content, and the META-INF directory, which contains certificates and signatures. Additionally, plain text resources like XML files or configuration files within the res directory, and certain asset files such as plain text documents or configuration files in the assets folder, might also exhibit low entropy. These areas generally have more predictable and uniform data compared to other sections of the APK file, such as compiled code or compressed resources. Conversely, areas of high entropy (yellow) could represent compressed or encrypted sections.

\begin{algorithm}[ht!]
    \SetAlgoLined
    \KwInput{List of Binary files}
    \KwOutput{Entropy plots }

    initialization\;
    
    $input\_files \leftarrow$ list all apk files \;

    \SetKwFunction{FCE}{entropy}
    \SetKwProg{Fn}{Function}{:}{}
    \Fn{\FCE{data, window\_size}}{
        $entropies \leftarrow []$;
        \For{$i \leftarrow 0$ \KwTo $length(data) - window\_size$ }{
            $window \leftarrow data[i : i + window\_size]$\;
            $counts \leftarrow$ frequency of each byte in window\;
            $probs \leftarrow counts / window\_size$\;
            $entropies$.append(entropy(probs, base=2))\;
        }
        \KwRet $entropies$\;
    }

    \SetKwFunction{PF}{process\_file}
    \SetKwProg{Fn}{Function}{:}{}
    \Fn{\PF{input\_file\_path, window\_size}}{
        $data \leftarrow$ read binary file data\;
        $byte\_array \leftarrow$ convert data to byte array\;
        $entropies \leftarrow$ entropy(byte\_array, window\_size)\;
        $normalized\_entropies \leftarrow$ normalize entropies to range 0-255\;
        $sqrt\_len \leftarrow$ ceil(sqrt(length(normalized\_entropies)))\;
        $padded\_entropies \leftarrow$ pad entropies with zeros to length $sqrt\_len^2$\;
        $entropy\_image \leftarrow$ reshape to 2D array of size $(sqrt\_len, sqrt\_len)$\;
       
    }

    \ForEach{file in input\_files}{
        process\_file(file)\;
    }

    \caption{Generate Entropy Plots from Binary Files}
\end{algorithm}

\subsection{Grayscale Data Mapping}
\begin{figure}[ht!]
	\centering
	\includegraphics[width=1\linewidth]{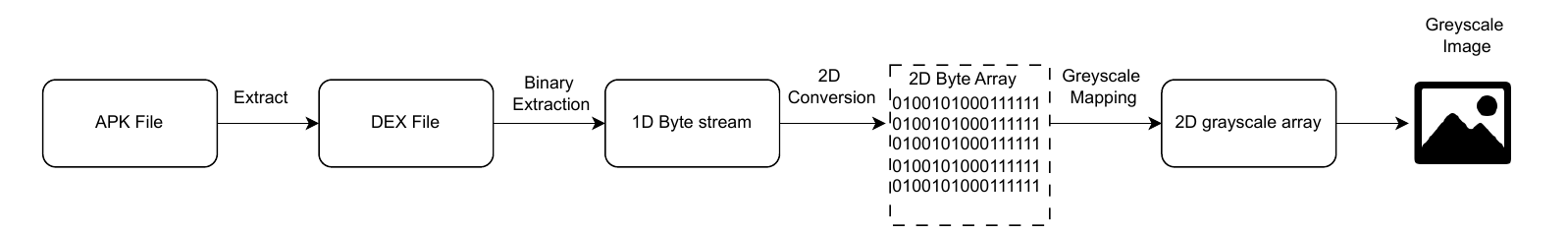}
	\caption{Steps of Grayscale Data Mapping}
	\label{figure:tundrod}
\end{figure}

The first step of Grayscale Data Mapping of an APK file is the reading of raw malware binary as a byte object as shown in Figure \ref{figure:tundrod}. For this purpose, the APK file is read by a binary file reader in read-binary mode. The mode is selected so that reading of the file occurs in the form of a stream of bytes without system modifications or interpretation so that the raw data remains unchanged. While the file is read, its contents are read into a byte object in its entirety. The byte object, in essence, a list of bytes, is an array of values between 0 and 255. Each value is mapped to a single byte of data copied from the APK file, without any processing, encompassing the raw binary information. The byte sequence is of the utmost significance since it is the first data to be mapped into grayscale in the following stages. With the raw preservation of the binary data of the APK file, this step ensures that the original form of the malware is preserved, with successful and correct data mapping.
Once the raw malware binary is imported into a byte object, the next step is to transform the byte object into a one-dimensional (1D) array of 8-bit unsigned integers. Each element in the array is between 0 and 255, with 0 being a black pixel and 255 being a white pixel. The 1D array is transformed into a two-dimensional (2D) array by determining the proper dimensions and reshaping the array. In cases where the length of the byte sequence does not form a perfect square, the data is padded with zeros to fill out the matrix. This padding is used to keep the matrix's equal dimensions, thus making it easy to create a uniform visual context.

\begin{figure}[ht!]
	\centering
	\includegraphics[width=0.6\linewidth]{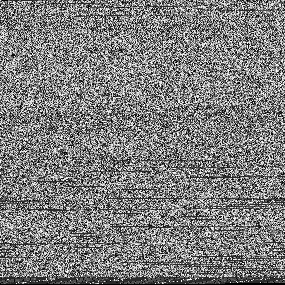}
	\caption{Grayscale Data of an application}
	\label{gray}
\end{figure}

The padded data is then reformatted into a two-dimensional array, a square array where every byte value is used to represent a pixel of the grayscale image. The array is then projected onto an image object through the use of the Python Imaging Library (PIL). The image is then resized into the desired size of 256x256 pixels, using the BICUBIC resampling filter \cite{gallagher2005detection} to maintain the visual integrity of the data presentation. The resizing process normalizes the image size, thus rendering it suitable for further analysis and comparison. A gray-level image, as shown in Figure \ref{gray}, is obtained from the raw binary data of a file and has a tightly packed pixel-value scale from black to white.
\textbf{Proposition:} No information is lost during grayscale mapping from binary files.

\textit{Justification:} Let \( B \) be a set of byte values where each \( b_i \in B \) is an 8-bit unsigned integer (0 to 255). The grayscale mapping function \( f: B \rightarrow G \) maps each byte value directly to a corresponding grayscale value \( g_i \) in the set \( G \) (0 to 255). This function is a one-to-one correspondence and reversible, allowing the original byte values to be fully recovered from the grayscale values. Therefore, the grayscale mapping process preserves all original data, ensuring no information is lost.

\begin{algorithm}[ht!]
\SetAlgoLined
\KwInput{List of Binary files}
\KwOutput{Grayscale images}
initialization\;

$input\_files \leftarrow$ list all files \;

\SetKwFunction{FGI}{create\_GrayImage}
\SetKwProg{Fn}{Function}{:}{}
\Fn{\FGI{input\_file, output\_dir, imgFormat, imageSize, resampling\_filter}}{
    $file\_content \leftarrow$ read binary file $input\_file$\;
    $data \leftarrow$ convert $file\_content$ to byte array of uint8\;
    $image\_size \leftarrow$ ceil(sqrt(length(data)))\;
    $padded\_data \leftarrow$ pad $data$ with zeros to length $image\_size^2$\;
    $image\_array \leftarrow$ reshape $padded\_data$ to 2D array of size $(image\_size, image\_size)$\;
    $image \leftarrow$ convert $image\_array$ to grayscale image\;
    $resized\_image \leftarrow$ resize $image$ to $(imageSize, imageSize)$ using $resampling\_filter$\;
    $output\_file\_path \leftarrow$ generate output path in $output\_dir$\;
    save $resized\_image$ to $output\_file\_path$\;
}

\ForEach{file in input\_files}{
    $output\_dir \leftarrow$ generate output directory\;
    \FGI{file, output\_dir, imgFormat, imageSize, resampling\_filter}\;
}

\caption{Generate Grayscale Images from Binary Files}
\end{algorithm}

The total time complexity of both the entropy analysis algorithm (Algorithm 1) and the grayscale image generation algorithm (Algorithm 2) is \( O(n + m^2) \), where \( n \) is the size of the input file and \( m \) is the dimension of the output image. For the entropy analysis algorithm, significant steps include calculating entropy values for sliding windows and saving the image. In contrast, for the grayscale image generation algorithm, key steps involve resizing the image and saving it. Both algorithms efficiently process the file data linearly but have quadratic time complexity due to image handling operations. However, this time complexity is much better than the dynamic analysis of a file, which often involves more computationally intensive operations such as runtime behavior monitoring, leading to significantly higher time complexity.

\section{Experiments and Results}\label{results}
All experiments has been implemented in Python using a Dell Precision 7810 workstation with 2x Intel Xeon (R) W-2145 comprising 8 cores, 64GB RAM, NVIDIA RTX 4090 GPU with 24GB VRAM, and Ubuntu OS. Materials used, preprocessing carried out, and performance achieved are discussed next.

\subsection{Dataset Characteristics}
\begin{figure}[ht!]
	\centering
\includegraphics[width=0.3\linewidth]{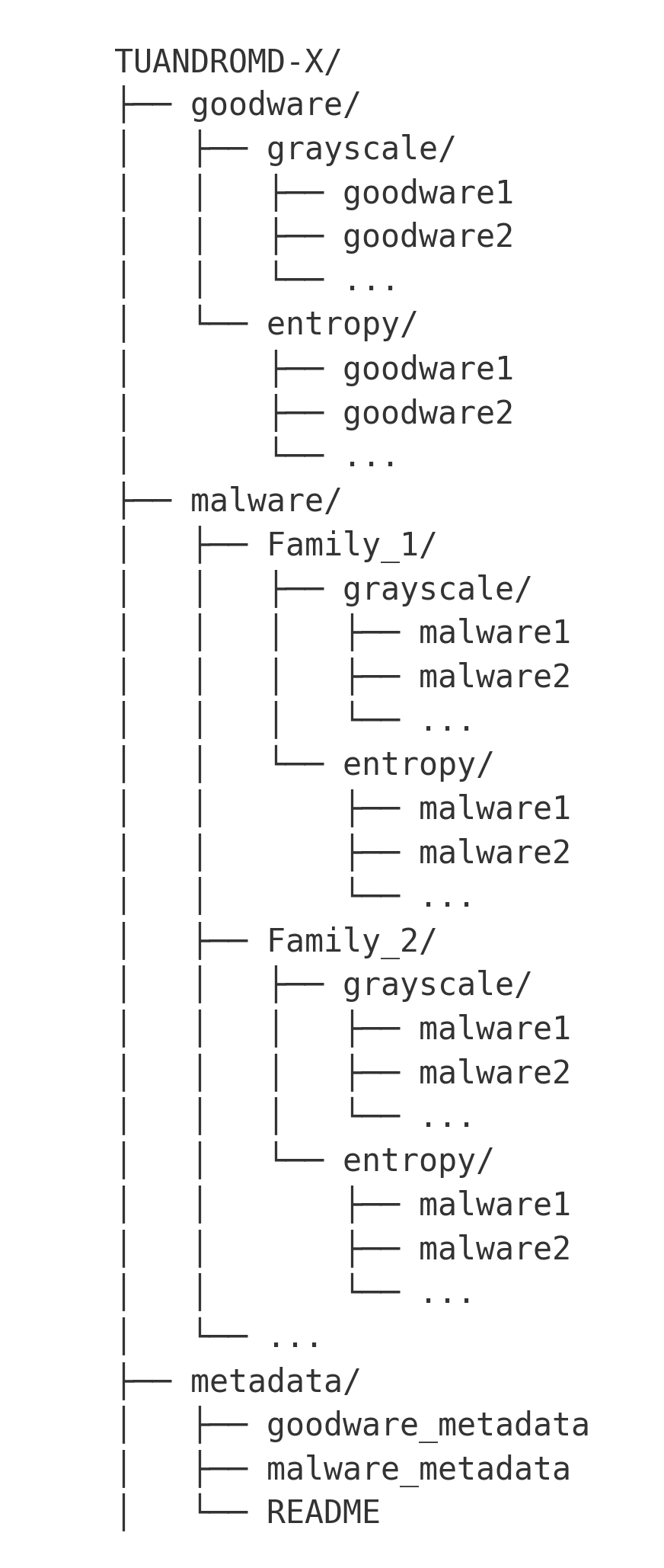}
	\caption{Directory structure of TUANDROMD-X}
	\label{DIRECT}
\end{figure}

\begin{figure}[ht!]
	\centering
	\includegraphics[width=0.8\linewidth]{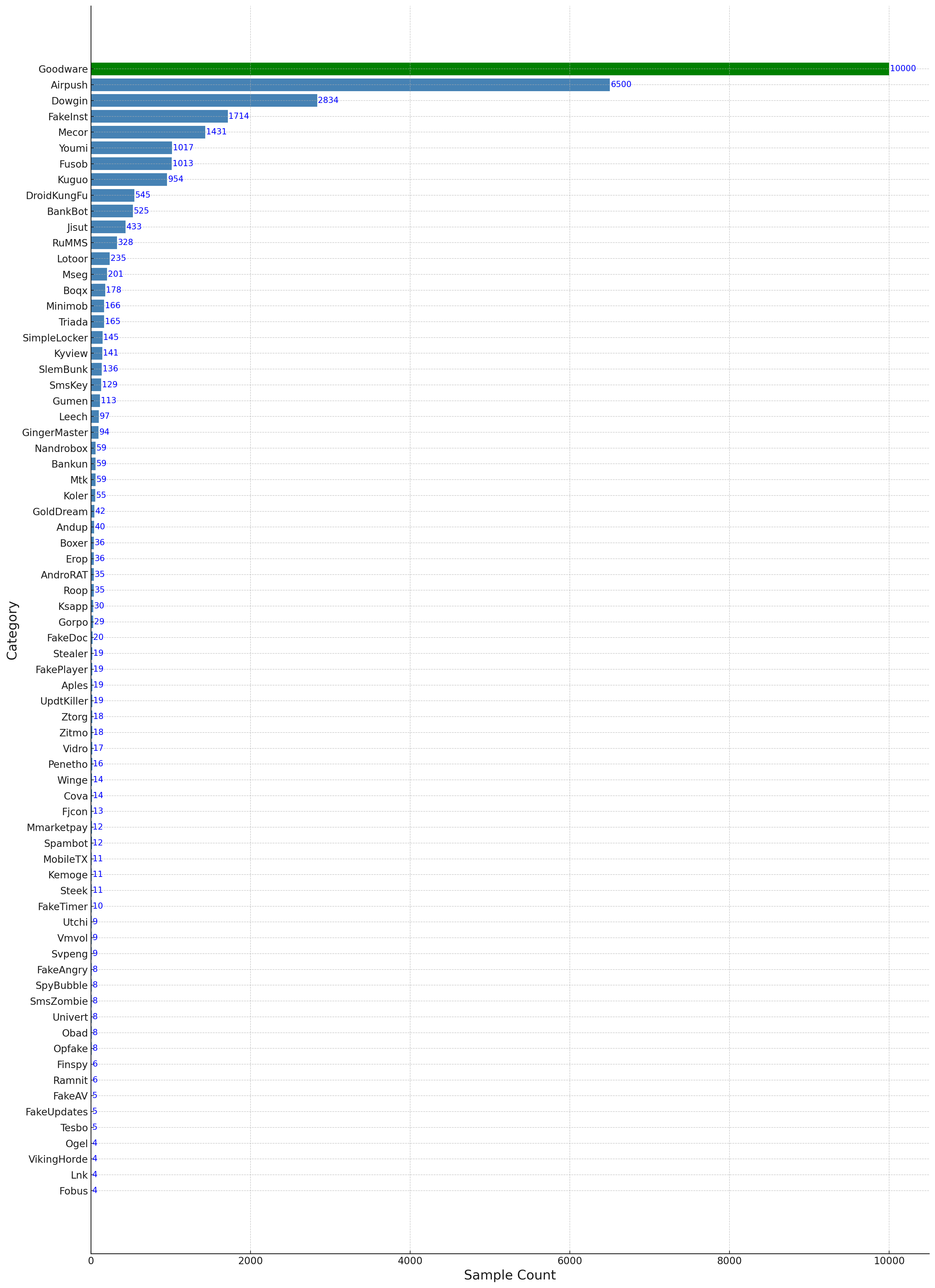}
	\caption{Class Distribution of TUANDROMD-X}
	\label{classd}
\end{figure}

\begin{figure}[ht!]
	\centering
	\includegraphics[width=0.8\linewidth]{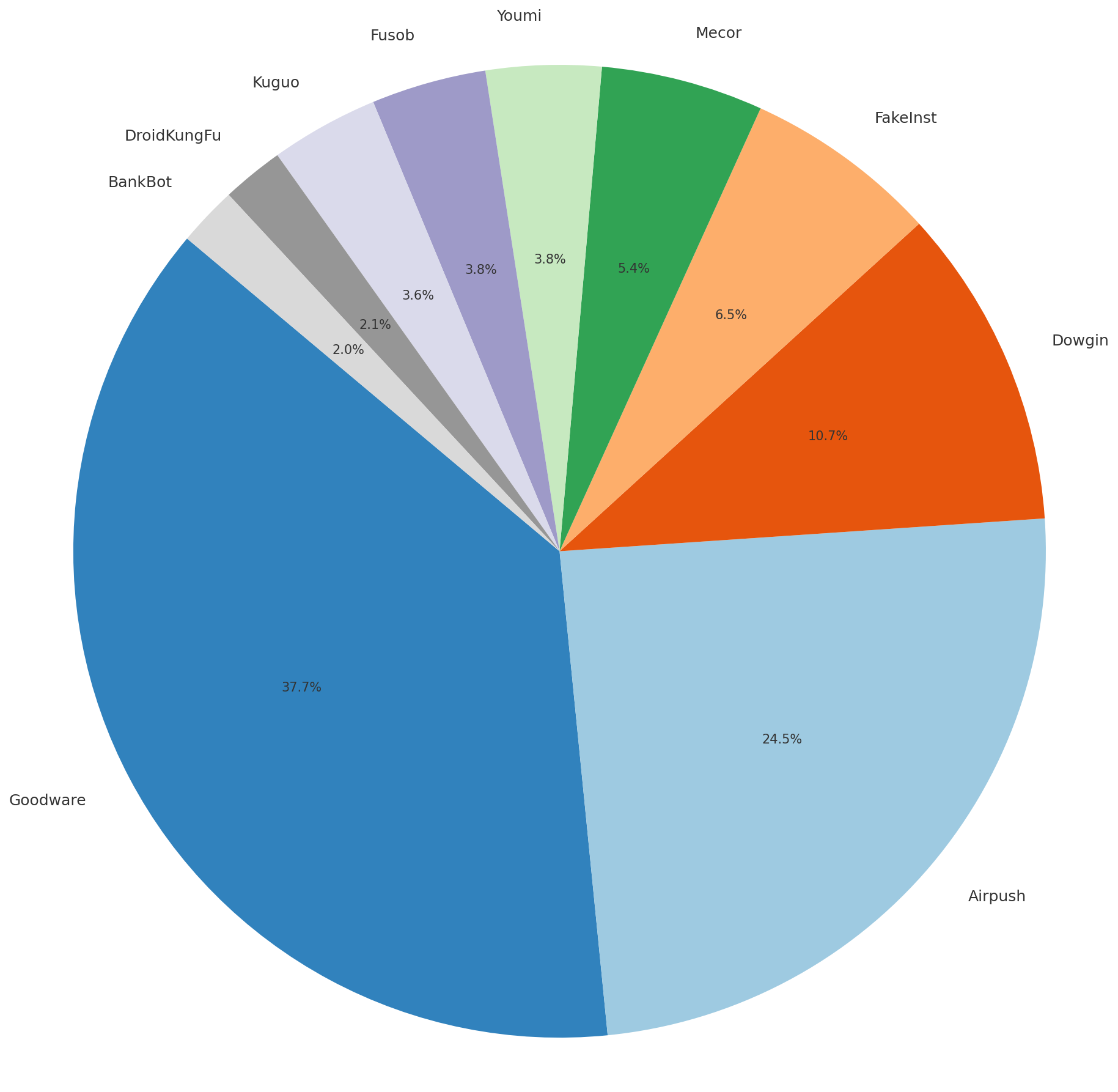}
	\caption{Top 10 categories of TUANDROMD-X}
	\label{top10}
\end{figure}

The TUANDROMD-X dataset consists of 30,000 instances of malware and goodware samples. It comprises 20,000 malware instances, and 10,000 goodware instances. The dataset is categorized into 72 distinct classes, with 71 classes dedicated to various types of malware and one class for goodware. The detailed dataset statistics are presented in Table~\ref{tab:dataset_statistics}. To provide a visual representation of the data distribution, Figure~\ref{classd} illustrates the overall class distribution, while Figure~\ref{top10} shows the top 10 categories within the dataset. 
The figure \ref{DIRECT} illustrates the directory structure, showcasing how the TUANDROMD-X dataset is organized.

\begin{table}[ht!]
\centering
\caption{Dataset Statistics}
\label{tab:dataset_statistics}
\begin{tabular}{lr}
\hline
\textbf{Characteristic} & \textbf{Count} \\
\hline
Total Instances & 30,000 \\
Malware Instances & 20,000 \\
Goodware Instances & 10,000 \\
\hline
Total Classes & 72 \\
Malware Classes & 71 \\
Goodware Class & 1 \\
\hline
\end{tabular}
\end{table}

\subsection{Performance Evaluation}
To evaluate the performance of our dataset's potential for malware classification, we employed state-of-the-art convolutional neural network (CNN) models \cite{yao2019review,dhillon2020convolutional}. We chose CNNs due to their proven success in image classification tasks across various fields, including cybersecurity. We implemented a strategic data partitioning scheme to ensure a thorough and unbiased assessment. Specifically, we divided the dataset into three distinct subsets: 70\% for training the models, 10\% for validation to fine-tune the models, and the remaining 20\% for testing to evaluate the final performance. By leveraging these advanced CNN architectures on our carefully partitioned dataset, we aim to establish a robust benchmark for malware classification performance.

\begin{table}[ht!]
\centering
\caption{Performance of CNN Models on Entropy Dataset}
\label{tab:cnn_performance}
\begin{tabular}{lcccc}
\hline
\textbf{Model} & \textbf{Accuracy (\%)} & \textbf{Precision(\%)} & \textbf{Recall(\%)} & \textbf{F1-Score (\%)} \\
\hline
ResNet-18 & 83.2 & 82.9 & 83.4 & 83.1 \\
ResNet-34 & 84.1 & 83.8 & 84.3 & 84.0 \\
ResNet-50 & 85.0 & 84.7 & 85.2 & 84.9 \\
DenseNet-121 & 79.8 & 79.5 & 80.0 & 79.7 \\
DenseNet-164 & 81.0 & 80.7 & 81.2 & 80.9 \\
\hline
\end{tabular}
\end{table}

\begin{table}[ht!]
\centering
\caption{Performance of CNN Models on Greyscale Image Dataset}
\label{tab:cnn_performance1}
\begin{tabular}{lcccc}
\hline
\textbf{Model} & \textbf{Accuracy (\%)} & \textbf{Precision (\%)} & \textbf{Recall (\%)} & \textbf{F1-Score (\%)} \\
\hline
ResNet-18 & 76.2 & 76.1 & 76.5 & 76.4 \\
ResNet-34 & 77.1 & 77.0 & 77.3 & 77.2 \\
ResNet-50 & 78.0 & 77.9 & 78.1 & 78.0 \\
DenseNet-121 & 79.8 & 79.7 & 79.2 & 79.1 \\
DenseNet-164 & 80.0 & 79.9 & 79.4 & 79.3 \\
\hline
\end{tabular}
\end{table}

\begin{figure}[ht!]
	\centering
	\includegraphics[width=1\linewidth]{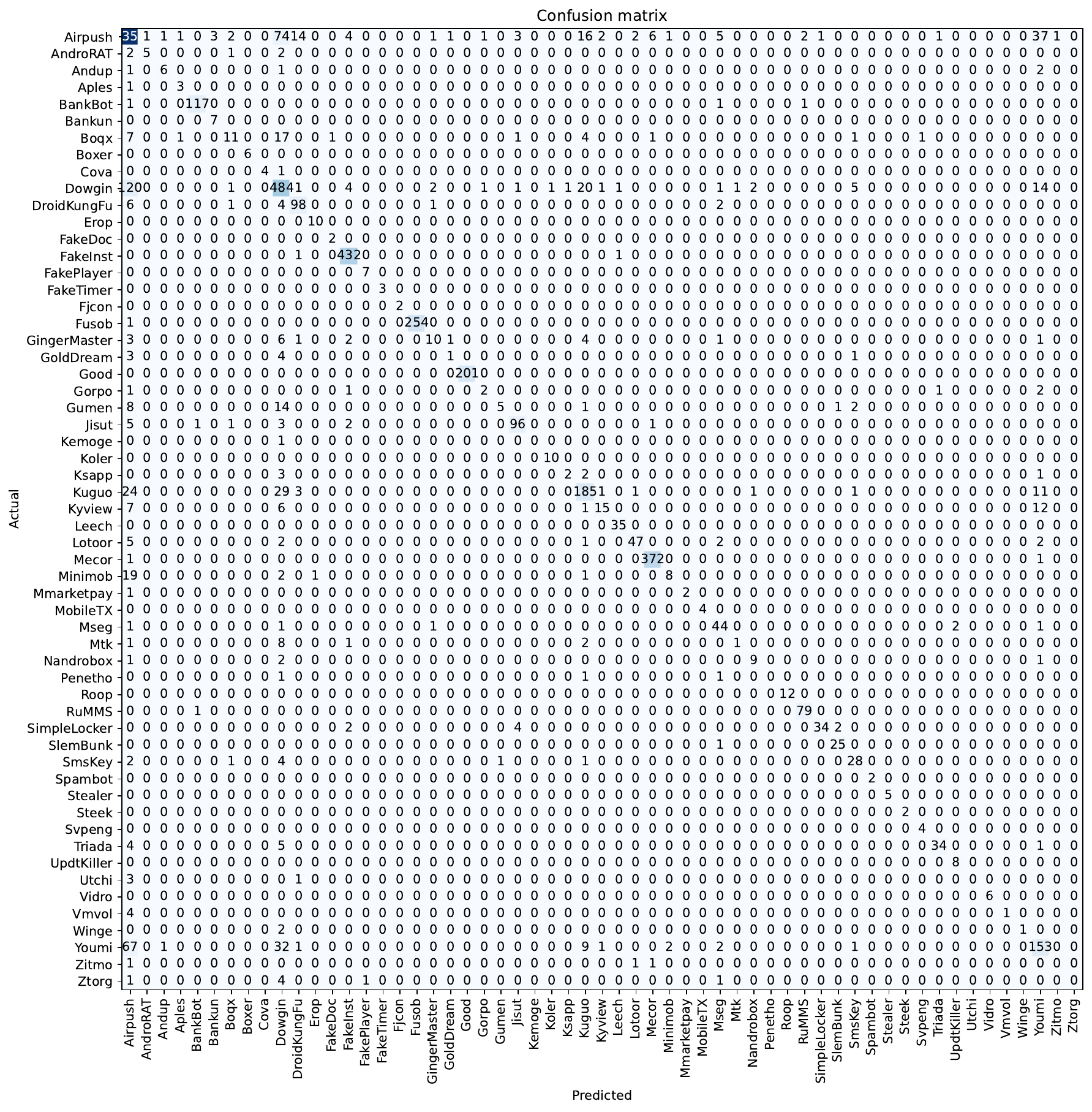}
	\caption{Confusion matrix of the Resnet 34 model on Entropy Dataset}
	\label{cmatrix}
\end{figure}

\begin{figure}[ht!]
	\centering
	\includegraphics[width=1\linewidth]{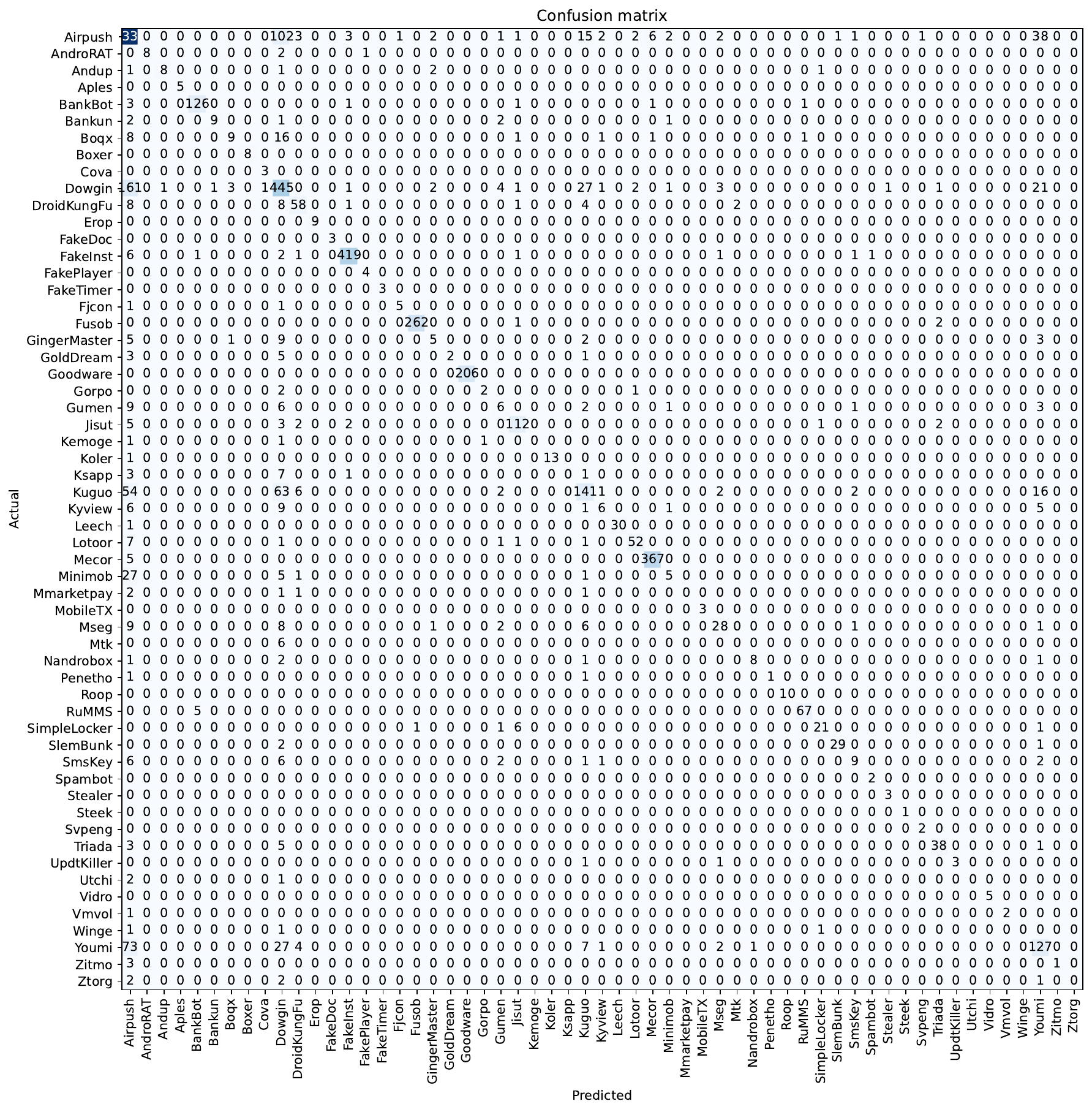}
	\caption{Confusion matrix of the Resnet 34 model on Greyscale Image Dataset}
	\label{cmatrix1}
\end{figure}
For performance evaluation, we considered five CNN models, namely ResNet-18, ResNet-34, ResNet-50, DenseNet-121, and DenseNet-164, and observed their performance on entropy-based and grayscale-based malware datasets. The results of these models on both datasets are reported in Table \ref{tab:cnn_performance} and \ref{tab:cnn_performance1}. On the entropy dataset, ResNet-18 achieved an accuracy of 83.2\%, while ResNet-34 and ResNet-50 showed improved performance with accuracies of 84.1\% and 85.0\%, respectively. DenseNet-121 and DenseNet-164 attained accuracies of 79.8\% and 81.0\%, respectively, on the same dataset. When evaluating the grayscale image dataset, the performance metrics were generally lower. ResNet-18, ResNet-34, and ResNet-50 achieved accuracies of 76.2\%, 77.1\%, and 78.0\%, respectively. DenseNet-121 maintained an accuracy of 79.8\%, similar to its performance on the entropy dataset, while DenseNet-164 again showed the highest performance with an accuracy of 80.0\%. These results indicate that the CNN models generally performed better on the entropy dataset compared to the grayscale image dataset. ResNet-50 had the highest overall performance on the entropy dataset, while DenseNet-164 performed best on both datasets.The confusion matrices for the ResNet-34 model, evaluated on both datasets, are presented in Figures \ref{cmatrix} and \ref{cmatrix1}. This evaluation highlights the importance of dataset characteristics in influencing model performance in the case of malware detection and underscores the need to choose the appropriate model architecture based on the specific nature of the data.

\begin{table}[ht!]
\centering
\caption{Performance of CNN Models on Entropy Dataset with Top 10 Classes}
\label{cnntop10}
\begin{tabular}{lcccc}
\hline
\textbf{Model} & \textbf{Accuracy (\%)} & \textbf{Precision(\%)} & \textbf{Recall(\%)} & \textbf{F1-Score (\%)} \\
\hline
ResNet-18 & 86.2 & 86.1 & 83.5 & 83.5 \\
ResNet-34 & 87.5 & 87.4 & 85.0 & 84.2 \\
ResNet-50 & 88.8 & 88.4 & 86.7 & 85.6 \\
DenseNet-121 & 86.0 & 85.9 & 83.2 & 82.8 \\
DenseNet-164 & 88.0 & 87.9 & 84.6 & 84.1 \\
\hline
\end{tabular}
\end{table}

\begin{table}[ht!]
\centering
\caption{Performance of CNN Models on Grayscale Dataset with Top 10 Classes}
\label{cnn1top10}
\begin{tabular}{lcccc}
\hline
\textbf{Model} & \textbf{Accuracy (\%)} & \textbf{Precision(\%)} & \textbf{Recall(\%)} & \textbf{F1-Score (\%)} \\
\hline
ResNet-18 & 82.3 & 82.2 & 81.5 & 81.4 \\
ResNet-34 & 83.2 & 83.1 & 82.3 & 82.2 \\
ResNet-50 & 84.0 & 83.9 & 83.1 & 83.0 \\
DenseNet-121 & 81.2 & 81.1 & 78.2 & 78.1 \\
DenseNet-164 & 83.0 & 82.9 & 79.4 & 79.3 \\
\hline
\end{tabular}
\end{table}

Tables \ref{cnntop10} and \ref{cnn1top10} show CNN models' performance on the Entropy Dataset and Grayscale Dataset, respectively, using the top 10 classes. For the Entropy Dataset, ResNet-50 achieved the highest accuracy at 88.8\%, while DenseNet-164 achieved accuracy at 88.0\%. In the Grayscale Dataset, ResNet-50 achieved the highest accuracy of 84.0\%, and DenseNet-164 performed with an accuracy of 83.0\%.

The performance of these models can be further enhanced by incorporating advanced hyperparameter tuning methods.  Data augmentation techniques can be integrated into our training pipeline to improve model generalization. Additionally, our dataset can be expanded and refined through targeted data collection efforts, which could involve increasing the overall dataset size and ensuring balanced representation across classes.

\subsection{Interpreting Model Focus on Malware Data Representations}
\begin{figure}[ht!]
    \centering
    \begin{minipage}[b]{0.45\linewidth}
        \centering
        \includegraphics[width=\linewidth]{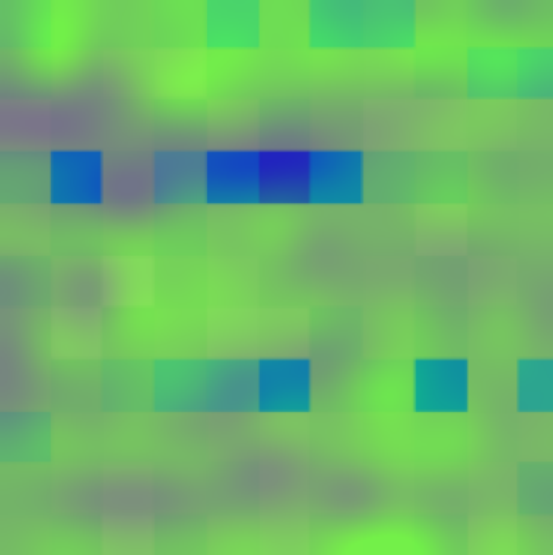}
        \textbf{(a) Entropy Data}
    \end{minipage}
    \hfill
    \begin{minipage}[b]{0.45\linewidth}
        \centering
        \includegraphics[width=\linewidth]{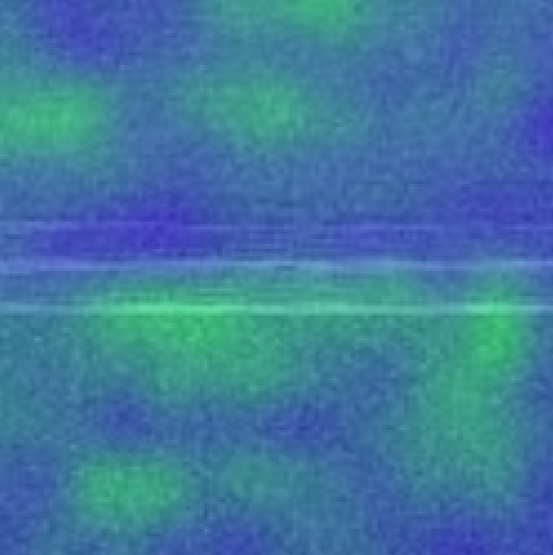}
        \textbf{(b) Grayscale Data}
    \end{minipage}
    \caption{ResNet-34 attention regions on entropy and grayscale data of Airpush malware}
    \label{cam}
\end{figure}

Figure \ref{cam} shows examples of the Class Activation Maps (CAMs) \cite{zhou2016learning} generated by a ResNet-34 model, on entropy and grayscale forms of Airpush malware. The colored areas in the figures refer to the areas of the malware data most related to the prediction of the model. The CAM of the grayscale figure explains the particular pixel patterns the model focuses on, and the CAM of the entropy figure highlights areas of varying entropy values most significant in differentiating the malware.
These visualizations are used to provide insights into the ResNet-34 model's decision-making, making it easier to see where in the data the model's classification is being affected the most. This is particularly useful for model verification, ensuring the model is focused on meaningful patterns, and identifying where model accuracy can be improved.

\subsection{Hyperparameters of CNN Models}
In this study, we evaluated various Convolutional Neural Network (CNN) models using a similar set of hyperparameters to ensure fair comparison. We employed a cyclical learning rate strategy \cite{smith2017cyclical}, oscillating between 0.001 and 0.01, to enhance model convergence and avoid local minima. Each model was trained for 30 epochs,and a batch size of 32 was used. The Adam optimizer was selected for its adaptive capabilities and robust performance across various tasks. To mitigate overfitting, dropout regularization with a rate of 0.5 was applied after each convolutional block.

\subsection{Limitations and Challenges}
One of the limitations of TUANDROMD-X is the problem of class imbalance. This is particularly true in case of the goodware samples and a few malware class samples, the reason being the difficulty to collect such samples and their public unavailability. In the case of such underrepresented malware classes, it is decided not to apply any augmentation technique as such augmentations might not accurately reflect the characteristics of the original raw malware. Artificially increasing the number of samples through image augmentation could potentially introduce artefacts or distortions that do not correspond to genuine malware behaviors or structures. Moreover, such augmentations might adversely affect the performance of machine learning models trained on the augmented data. By maintaining the original distribution even though imbalanced, we aim to preserve the authenticity and integrity of the malware representations in TUANDROMD-X. This approach ensures that our analysis and subsequent model training is based on genuine, unaltered malware samples, providing a more realistic assessment of model performance in real-world scenarios where data imbalance is common. Additionally, the difficulty in collecting samples from underrepresented malware classes poses a significant challenge. For malware types that are not publicly available, data collection may require the setup of advanced honeynets, which can be costly as well as resource-intensive. One solution to address the scarcity of samples from rare malware classes is the collaborative efforts within the cybersecurity community to facilitate the safer sharing of malware samples. Also, developing more sophisticated and cost-effective honeypot technologies could aid in capturing a wider variety of malware samples. Although, Shannon entropy estimation is helpful in the effective detection of malware, there are other entropy measures that need to be explored to help discriminate the minor variations between the goodware and some of the malware instances.

\section{Conclusion}\label{conclusion}
In this study, we introduce TUANDROMD-X, a novel malware database developed using static analysis methods. TUANDROMD-X comprises of two distinct representations of malware samples: one based on entropy values and another based on grayscale image representations. The primary significance of TUANDROMD-X lies in its efficient generation process, which is considerably faster than databases relying on dynamic analysis techniques. By leveraging static analysis, TUANDROMD-X offers researchers a valuable resource for rapid malware classification and analysis. The dataset is evaluated using various Convolutional Neural Network (CNN) models, demonstrating its utility in training and benchmarking machine learning algorithms for malware detection. 


\section*{Declarations}
\subsection*{Conflict of interest}
Conflict of Interest:  On behalf of all authors, the corresponding author states that there is no conflict of interest.
 

\subsection*{Ethics approval and consent to participate}
Not applicable
\subsection*{Availability of data and material}
Made available on request
\subsection*{Consent for publication}
On behalf of all the authors, ``I, the Corresponding Author, declare that this manuscript is original, has not been published before, and is not currently being considered for publication elsewhere.''
\subsection*{Funding}
Not applicable


\bibliography{sample-base}


\end{document}